\documentstyle[11pt]{article}
\begin{document}
\title{General Gauge Field Theory }
\author{{Ning Wu}\\
{\small CCAST (World Lab), P.O.Box 8730, Beijing 100080, P.R.China}\\
{\small and}\\ 
{\small Division 1, Institute of High Energy Physics, P.O.Box 918-1, 
Beijing 100039, P.R.China}}
\maketitle
\vskip 0.8in

\vskip 0.8in

\noindent
[Abstract] In this paper, we will construct a gauge field model,  in  which the masses of  gauge 
fields are non-zero and the local gauge symmetry is strictly preserved.  A $SU(N)$ gauge field 
model is discussed in details in this paper. In the limit $\alpha \longrightarrow 0$ or $\alpha 
\longrightarrow \infty$, the gauge field model discussed in this paper will return to Yang-Mills 
gauge field model.This theory could be regarded as theoretical development of Yang-Mills gauge
field theory.  \\

\Roman{section}
\section{Introduction}

~~~~In 1954, Yang and Mills founded the non-Abel gauge field theory \lbrack 1 \rbrack . Since then, 
the gauge field theory has developed greatly, and has been widely applied to elementary particle 
theory. Now, we believe that, four kinds of fundamental interactions in nature are all gauge 
interactions and can be described by gauge field theories. It is generally believed that the principle of 
local gauge invariance should play a fundamental role in interaction theory. In the sixties, the gauge 
field theory was applied to electro-weak interactions, and the $SU(2)_L \times U(1)_Y$ unified 
electro-weak gauge theory was founded \lbrack 2-4 \rbrack. In the seventies, gauge field theory was 
applied to strong interaction, and the $SU(3)_c$ quantum chromodynamics was founded.  \\

~~~~But, we know that, if the local gauge symmetry is strictly preserved in Yang-Mills gauge field 
theory, the masses of gauge fields should be zero. However in the forties, physicists have realized  
that the intermediate bosons, which transmit weak interaction, should have very big masses \lbrack 5 
\rbrack. A possible way to solve this problem is to introduce the concept of spontaneously symmetry 
breaking \lbrack 6-8 \rbrack and Higgs mechanism \lbrack 9-13 \rbrack. Higgs mechanism plays an 
important role in the standard model. The gauge bosons $W^{\pm}$ and $Z^0$, which are 
predicted 
by the standard model theoretically, are discovered by experiment in the eighties, but the Higgs 
particle, which is necessary for the standard model, is not found until now. Does Higgs particle exist 
in nature? If there were no Higgs particle, how could the intermediate bosons $W^{\pm}$ and $Z^0$ 
obtain masses? In the theoretical point of view, can we construct a gauge field model, in which we 
need no Higgs mechanism, but the  masses of  gauge bosons are non-zero. \\

~~~~In this paper, we will construct a gauge field model,  which has strict local gauge symmetry 
and massive gauge fields. In order to do this, two sets of gauge fields are needed. In this paper, 
We will give the lagrangian of the model first, then prove the gauge 
symmetry of the model and deduce the conserved charges which correspond to the gauge symmetry . 
After that, we will construct the eigenvectors of mass matrix, and deduce the equations of motion of 
all fields. The case that matter fields are boson fields is also discussed in this paper. A more 
general gauge field model is given in chapter eight. After that, we discuss the Yang-Mills limit of the 
present theory.  Finally, we present some discussions. \\

\section{The Lagrangian of The Model}

~~~~ For the sake of generality, we let the gauge group be $SU(N)$ group, and for the sake of 
simplicity, we select fermion fields as matter fields. Suppose that $N$ fermoin fields $\psi _l 
(x)~(l=1,2, \ldots ,N)$ form a multiplet of matter fields. let\\
$$
\psi (x) =\left ( 
\begin{array}{c}
\psi_1 (x) \\
\psi_2 (x) \\
\vdots \\
\psi_N (x)
\end{array}
\right ) 
\eqno{(2.1)} 
$$
$\psi (x)$ is a N-component vector. All $\psi (x)$ form a space of fundamental representation 
of  $SU(N)$ group. In this representative space, the representative matrices of the generators of 
$SU(N)$ group are denoted by $T_i ~(i=1,2, \ldots, N^2-1)$. They satisfy:
$$
\lbrack T_i ~ ,~ T_j \rbrack = i f_{ijk} T_k
\eqno{(2.2)} 
$$
$$
Tr( T_i  T_j ) = \delta_{ij} K.
\eqno{(2.3)} 
$$
where $f_{ijk}$ are structure constants of $SU(N)$ group, K is a constant 
which is independent of indices $i$ and $j$ but depends on the representation 
of the group. Generators $T_i$ are  Hermit and traceless:
$$
T_i ^{\dag}=T_i
\eqno{(2.4)} 
$$
$$
Tr  T_i = 0.
\eqno{(2.5)} 
$$
The representative matrix of a general element of the $SU(N)$ group can be 
written as:
$$
U=e^{-i \alpha ^i T_i}
\eqno{(2.6)} 
$$
with $\alpha ^i$ the real group parameters. $U$ is a unitary $N \times N$ matrix
$$
U^{\dag} U = 1 = U U^{\dag}
\eqno{(2.7)} 
$$
In global gauge transformation, all $\alpha ^i$ are independent of space-time coordinates, where in 
local gauge transformation, $\alpha ^i$ are functions of space-time coordinates. \\

~~~~ Two kinds of vector fields $A_{\mu}(x)$ and $B_{\mu}(x)$ will be introduced in this paper. 
$A_{\mu}(x)$ and $B_{\mu}(x)$ are vectors in the adjoint representation of $SU(N)$ group. They 
can be expressed as linear combinations of generators :
$$
A_{\mu}(x) = A_{\mu} ^i (x) T_i
\eqno{(2.8a)} 
$$
$$
B_{\mu}(x) = B_{\mu} ^i (x) T_i.
\eqno{(2.8b)}
$$
$A_{\mu}^i (x)$ and $B_{\mu}^i (x)$ are component fields of gauge fields $A_{\mu}(x)$ and 
$B_{\mu}(x)$ respectively. \\

~~~~ Corresponds to two kinds of gauge fields, there are two kinds of gauge covariant derivatives 
in the theory:
$$
D_{\mu} = \partial _{\mu} - ig A_{\mu}
\eqno{(2.9a)} 
$$
$$
D_{b \mu} = \partial _{\mu} + i \alpha g B_{\mu}.
\eqno{(2.9b)} 
$$
The strengths of gauge fields $A_{\mu}(x)$ and $B_{\mu}(x)$ are defined as 
$$
\begin{array}{ccl}
A_{\mu \nu} & = & \frac{1}{-i g} \lbrack D_{\mu} ~,~ D_{\nu} \rbrack
\\
& = & \partial _{\mu} A_{\nu} - \partial _{\nu} A_{\mu}
- i g \lbrack A_{\mu} ~,~ A_{\nu} \rbrack
\end{array}
\eqno{(2.10a)} 
$$
$$
\begin{array}{ccl}
B_{\mu \nu} &=& \frac{1}{i \alpha g} \lbrack D_{b \mu} ~,~ D_{b \nu} \rbrack
\\
& = & \partial _{\mu} B_{\nu} - \partial _{\nu} B_{\mu}
+ i \alpha g \lbrack B_{\mu} ~,~ B_{\nu} \rbrack. 
\end{array}
\eqno{(2.10b)} 
$$
respectively. Similarly, $A_{\mu \nu}$ and $B_{\mu \nu}$ can also be expressed as linear 
combinations of generators:
$$
A_{\mu \nu} = A_{\mu \nu}^i T_i
\eqno{(2.11a)} 
$$
$$
B_{\mu \nu}= B_{\mu \nu}^i T_i. 
\eqno{(2.11b)} 
$$
Using relations (2.2) and (2.10a,b), we could obtain
$$
A_{\mu \nu}^i = \partial _{\mu} A_{\nu}^i - \partial _{\nu} A_{\mu}^i
+g f^{ijk} A_{\mu}^j    A_{\nu}^k
\eqno{(2.12a)} 
$$
$$
B_{\mu \nu}^i = \partial _{\mu} B_{\nu}^i - \partial _{\nu} B_{\mu}^i
- \alpha g f^{ijk} B_{\mu}^j    B_{\nu}^k .
\eqno{(2.12b)} 
$$

~~~~ The lagrangian density of the model is
$$
\begin{array}{ccl}
\cal L &= &- \overline{\psi}(\gamma ^{\mu} D_{\mu} +m) \psi 
-\frac{1}{4K} Tr( A^{\mu \nu} A_{\mu \nu} )
-\frac{1}{4K} Tr( B^{\mu \nu} B_{\mu \nu} ) \\
&&-\frac{\mu ^2}{2K ( 1+ \alpha ^2)} 
Tr \left \lbrack (A^{\mu}+\alpha B^{\mu})( A_{\mu}+\alpha B_{\mu} ) 
\right \rbrack
\end{array}
\eqno{(2.13)} 
$$
where $\alpha$ is a constant. In this paper, the space-time  metric is selected as $ \eta _{\mu \nu} = 
diag (-1,1,1,1)$, $(\mu ,\nu =0,1,2,3)$. According to relation (2.3), the above lagrangian density 
${\cal L}$ can be rewritten as:
$$
\begin{array}{ccl}
\cal L &= &- \overline{\psi} \lbrack \gamma ^{\mu} ( \partial _{\mu}
 - i g A^i_{\mu} T_i) +m \rbrack \psi 
-\frac{1}{4}  A^{i \mu \nu} A^i_{\mu \nu} 
-\frac{1}{4}  B^{i \mu \nu} B^i_{\mu \nu} \\
&&-\frac{\mu ^2}{2 ( 1+ \alpha ^2)} 
(A^{i \mu}+\alpha B^{i \mu})( A^i_{\mu}+\alpha B^i_{\mu} ) 
\end{array}
\eqno{(2.14)} 
$$
It is easy to see that, except for the mass term and the term concerned with gauge field $B_{\mu}$, 
the above lagrangian density is the same as that of Yang-Mills theory. \\

\section{Global Gauge Symmetry and Conserved Charges}

~~~~ Now we discuss the gauge symmetry of the lagrangian density ${\cal L}$. First, we discuss 
the global gauge symmetry and the corresponding conserved charges. In global gauge 
transformation, 
the matter field $\psi$ transforms as:
$$
\psi \longrightarrow \psi ' = U \psi ,
\eqno{(3.1)} 
$$
where $U$ is a $N \times N$ transformation matrix which is defined by eq(2.6). $U$ is independent 
of space-time coordinates. That is 
$$
\partial _{\mu} U = 0. 
\eqno{(3.2)} 
$$
The corresponding global gauge transformations of gauge fields $A_{\mu}$ and $B_{\mu}$ are
$$
A_{\mu} \longrightarrow U A_{\mu} U^{\dag}
\eqno{(3.3a)} 
$$
$$
B_{\mu} \longrightarrow U B_{\mu} U^{\dag}
\eqno{(3.3b)} 
$$
respectively. It is easy to prove that 
$$
D_{\mu} \longrightarrow U D_{\mu} U^{\dag}
\eqno{(3.4a)} 
$$
$$
D_{b \mu} \longrightarrow U D_{b \mu} U^{\dag}
\eqno{(3.4b)} 
$$
$$
A_{\mu \nu} \longrightarrow U A_{\mu \nu} U^{\dag}
\eqno{(3.5a)} 
$$
$$
B_{\mu \nu} \longrightarrow U B_{\mu \nu} U^{\dag}
\eqno{(3.5b)} 
$$
Using all the above transformation relations, we can prove that every term in eq(2.13) is gauge 
invariant. So, the whole lagrangian density has global gauge symmetry.  \\

~~~~ Let $\alpha ^i$ in eq(2.6) be the first order infinitesimal parameters, then the transformation 
matrix $U$ can be rewritten as:
$$
U \approx 1 - i \alpha ^i T^i .
\eqno{(3.6)} 
$$
The first order infinitesimal changes of fields $\psi, ~\overline{\psi}, ~A_{\mu}$ and $B_{\mu}$ 
are
$$
\delta \psi = - i \alpha ^i T^i \psi
\eqno{(3.7a)} 
$$
$$
\delta  \overline {\psi} =  i \alpha ^i \overline{\psi} T^i
\eqno{(3.7b)} 
$$
$$
\delta A_{\mu} =  \alpha ^i f^{ijk} A^j_{\mu} T^k 
\eqno{(3.8a)} 
$$
$$
\delta B_{\mu} =  \alpha ^i f^{ijk} B^j_{\mu} T^k 
\eqno{(3.8b)} 
$$
respectively. From eqs(3.8a,b) and eqs(2.8a,b), we can obtain
$$
\delta A_{\mu}^k =  \alpha ^i f^{ijk} A^j_{\mu}
\eqno{(3.9a)} 
$$
$$
\delta B_{\mu}^k =  \alpha ^i f^{ijk} B^j_{\mu}.
\eqno{(3.9b)} 
$$

~~~~ The first order variation of the lagrangian density is 
$$
\begin{array}{ccl}
\delta {\cal L} & = &
\partial _{\mu} \left (
\frac{ \partial {\cal L}}{\partial \partial _{\mu} \psi} \delta \psi 
+  \delta \overline{\psi}  \frac{ \partial {\cal L}}{\partial \partial _{\mu} \overline{\psi}}
+  \frac{ \partial {\cal L}}{\partial \partial _{\mu} A_{\nu}^k } \delta A_{\nu}^k
+  \frac{ \partial {\cal L}}{\partial \partial _{\mu} B_{\nu}^k } \delta B_{\nu}^k
\right )
\\
& = &
\alpha ^i \partial ^{\mu} J^i_{\mu},
\end{array}
\eqno{(3.10)} 
$$
where
$$
J^i_{\mu} = 
i \overline{\psi} \gamma _{\mu} T^i \psi
- f^{ijk} A^{j \nu} A^k _{\mu \nu}
- f^{ijk} B^{j \nu} B^k _{\mu \nu}.
\eqno{(3.11)} 
$$
The conserved current can also be written as
$$
\begin{array}{ccl}
J_{\mu} & = & 
i \overline{\psi} \gamma _{\mu} T^i \psi T^i
+ i \lbrack  A^{ \nu}  ~,~  A _{\mu \nu} \rbrack 
+ i \lbrack  B^{ \nu}  ~,~  B _{\mu \nu} \rbrack 
\\
& = & J^i _{\mu} T^i . 
\end{array}
\eqno{(3.12)} 
$$
Because the lagrangian density ${\cal L}$ has global gauge symmetry, the currents $J^i_{\mu}$ are 
conserved currents. They satisfy 
$$
\partial ^{\mu} J^i_{\mu} = 0 .
\eqno{(3.13)}
$$

~~~~ The corresponding conserved charges are
$$
\begin{array}{ccl}
Q^i & = &  \int d^3 x  J^{i 0}
\\
& = &  \int d^3 x (
\psi ^{\dag} T^i \psi 
+  \lbrack  A_j   ~,~  A ^{j 0} \rbrack  ^i
+  \lbrack  B_j   ~,~  B ^{j 0} \rbrack  ^i
) .
\end{array}
\eqno{(3.14)} 
$$
After quantization, $Q^i$ are the generators of gauge transformation. At the same time, we note 
that, no matter what is the value of parameter $\alpha$, gauge fields $A_{\mu}$ and $B_{\mu}$ 
contribute the same terms to the conserved currents and conserved charges. \\

\section{Local Gauge Symmetry}

~~~~ If $U$ in eq(3.1) depends on space-time coordinates, the transformation of eq(3.1) is a local 
gauge transformation. In this case,
$$
\partial _{\mu} U \not= 0 ~,~~ \partial _{\mu} \alpha^i \not= 0
\eqno{(4.1)}
$$
The corresponding transformations of gauge fields $A_{\mu}$ and $B_{\mu}$ are 
$$
A_{\mu} \longrightarrow U A_{\mu} U^{\dag}
-\frac{1}{ig}U \partial _{\mu}U^{\dag}
\eqno{(4.2a)} 
$$
$$
B_{\mu} \longrightarrow U B_{\mu} U^{\dag}
+\frac{1}{i \alpha g}U \partial _{\mu}U^{\dag}
\eqno{(4.2b)} 
$$
respectively.

~~~~ Using above relations, it is easy to prove that
$$
D_{\mu} \longrightarrow U D_{\mu} U^{\dag}
\eqno{(4.3a)} 
$$
$$
D_{b \mu} \longrightarrow U D_{b \mu} U^{\dag}.
\eqno{(4.3b)} 
$$
Therefore,
$$
A_{\mu \nu} \longrightarrow U A_{\mu \nu} U^{\dag}
\eqno{(4.4a)} 
$$
$$
B_{\mu \nu} \longrightarrow U B_{\mu \nu} U^{\dag}
\eqno{(4.4b)} 
$$
$$
D_{\mu} \psi  \longrightarrow U D_{\mu} \psi
\eqno{(4.5)} 
$$
$$
A_{\mu} + \alpha B_{\mu} \longrightarrow U  (A_{\mu} + \alpha B_{\mu}  )  U^{\dag}
\eqno{(4.6)} 
$$
Using all these transformation relations, we could prove that the lagrangian density ${\cal L}$ 
defined by eq(2.13) is invariant under local gauge transformations. Therefore the model has strict 
local gauge symmetry.  \\

\section{The Masses of Gauge Fields}

~~~~ If we select $A_{\mu}$ and $B_{\mu}$ as basis, the mass matrix on this basis is
$$
M  = \frac{1}{1+\alpha ^2} \left ( 
\begin{array}{cc}
\mu ^2  &  \alpha \mu^2  \\
\alpha \mu ^2  &  \alpha^2 \mu^2
\end{array}
\right ) ,
\eqno{(5.1)} 
$$
and the mass term of gauge fields can be written as:
$$
( A_{\mu} ~,~ B_{\mu} )  M  \left ( 
\begin{array}{c}
A_{\mu}  \\
B_{\mu}
\end{array}
\right ) .
\eqno{(5.2)} 
$$

~~~~ The particles generated from gauge interaction should be eigenvectors of mass matrix and the 
corresponding masses of these particles should be eigenvalues of mass matrix. $M$ has two 
eigenvalues, they are
$$
m^2_1 = \mu ^2 ~~,~~  m^2_2 = 0.
\eqno{(5.3)} 
$$
The corresponding eigenvectors are
$$
\frac{1}{\sqrt{1+\alpha ^2}} \left ( 
\begin{array}{c}
1  \\  \alpha
\end{array}
\right )
~~ , ~~
\frac{1}{\sqrt{1+\alpha ^2}} \left ( 
\begin{array}{c}
- \alpha  \\  1
\end{array}
\right ) ,
\eqno{(5.4)} 
$$
respectively. If we define
$$
{\rm  cos} \theta =  \frac{1}{\sqrt{1+\alpha ^2}}
  ~~,~~  
{\rm  sin} \theta =  \frac{\alpha}{\sqrt{1+\alpha ^2}} .
\eqno{(5.5)} 
$$
Then, eq(5.4) changes into
$$
\left ( 
\begin{array}{c}
{\rm cos } \theta  \\  {\rm sin} \theta
\end{array}
\right)
~~ , ~~
\left ( 
\begin{array}{c}
- {\rm sin } \theta  \\  {\rm cos} \theta
\end{array}
\right) ,
\eqno{(5.6)} 
$$

~~~~ Define
$$
C_{\mu}={\rm cos}\theta A_{\mu}+{\rm sin}\theta B_{\mu}
\eqno{(5.7a)} 
$$
$$
F_{\mu}=-{\rm sin}\theta A_{\mu}+{\rm cos}\theta B_{\mu}.
\eqno{(5.7b)} 
$$
It is easy to know that $C_{\mu}$ and $F_{\mu}$ are eigenstates of mass matrix, they describe 
those 
particles generated from gauge interaction. The inverse transformations of (5.7a,b) are 
$$
A_{\mu}= {\rm cos}\theta C_{\mu} - {\rm sin}\theta F_{\mu}
\eqno{(5.8a)} 
$$
$$
B_{\mu}= {\rm sin}\theta C_{\mu}+{\rm cos}\theta F_{\mu}.
\eqno{(5.8b)} 
$$
Then the lagrangian density ${\cal L}$ given by (2.14) changes into:
$$
{\cal L} = {\cal L}_0 + {\cal L}_I ,
\eqno{(5.9)}
$$
where
$$
{\cal L}_0= - \overline{\psi}(\gamma ^{\mu} \partial _{\mu} +m) \psi 
-\frac{1}{4} C^{i \mu \nu}_0 C^i_{0 \mu \nu} 
-\frac{1}{4} F^{i \mu \nu}_0 F^i_{0 \mu \nu}
-\frac{\mu ^2}{2} C^{i \mu} C^i_{\mu}.
\eqno{(5.10a)} 
$$
$$
\begin{array}{ccl}
{\cal L}_I & = & i g \overline{\psi} \gamma ^{\mu} ( {\rm cos}\theta C_{\mu}
 - {\rm sin}\theta F_{\mu} )  \psi  \\
&& 
- \frac{{\rm cos}2 \theta}{2 {\rm cos} \theta} g f^{ijk}C_0^{i \mu \nu} C^j_{\mu} C^k_{\nu}
+\frac{{\rm sin} \theta}{2 } g f^{ijk}F_0^{i \mu \nu} F^j_{\mu} F^k_{\nu}  \\
&&
+\frac{{\rm sin} \theta}{2 } g f^{ijk}F_0^{i \mu \nu} C^j_{\mu} C^k_{\nu}  
+ g {\rm sin} \theta f^{ijk} C_0^{i \mu \nu} C^j_{\mu} F^k_{\nu}  \\
&&
- \frac{1 - \frac{3}{4}{\rm sin}^2 2 \theta}{4 {\rm cos}^2 \theta} g^2 
f^{ijk} f^{ilm}  C^j_{\mu} C^k_{\nu} C^{l \mu} C^{m \nu}  \\
&&
- \frac{{\rm sin}^2  \theta}{4} g^2 f^{ijk} f^{ilm} F^j_{\mu} F^k_{\nu} F^{l \mu} F^{m \nu}
+ g^2 {\rm tg} \theta {\rm cos} 2 \theta f^{ijk} f^{ilm} C^j_{\mu} C^k_{\nu} C^{l \mu} F^{m \nu}
\\
&&
- \frac{{\rm sin}^2  \theta}{2} g^2 f^{ijk} f^{ilm} ( C^j_{\mu} C^k_{\nu} F^{l \mu} F^{m \nu}
+ C^j_{\mu} F^k_{\nu} F^{l \mu} C^{m \nu}+ C^j_{\mu} F^k_{\nu} C^{l \mu} F^{m \nu}) .
\end{array}
\eqno{(5.10b)} 
$$
In the above relations, we have used the following notations:
$$
C_{0 \mu \nu}^i = \partial _{\mu} C_{\nu}^i - \partial _{\nu} C_{\mu}^i
\eqno{(5.11a)} 
$$
$$
F_{0 \mu \nu}^i = \partial _{\mu} F_{\nu}^i - \partial _{\nu} F_{\mu}^i
\eqno{(5.11b)} 
$$

~~~~ From eq(5.10a), it is easy to see that the mass of gauge field $C_{\mu}$ is $\mu$ and the mass 
of gauge field $F_{\mu}$ is zero. That is
$$
m_c = \mu ~~,~~ m_F = 0 .
\eqno{(5.12)}
$$
Please note that, up to now, the gauge symmetry is strictly preserved. Therefore, without Higgs 
mechanism, gauge fields can have non-zero masses. Strict gauge symmetry does not mean that the  
gauge fields are all massless.  \\

\section{Equation of Motion}

~~~~ The Euler-Lagrange equation of motion for fermion field can be deduced from eq(5.9):
$$
\lbrack \gamma ^{\mu} ( \partial _{\mu} - i g {\rm cos}\theta C_{\mu}
+ i g {\rm sin}\theta F_{\mu} ) +m \rbrack  \psi = 0 .
\eqno{(6.1)}
$$
If we deduce the Euler-Lagrange equations of motion for gauge fields from eq(5.9), we will obtain 
very complicated expressions. For the sake of simplicity, we deduce the equations of motion for 
gauge fields from eq(2.13). In this case, the equations of motion for gauge fields $A_{\mu}$ and 
$B_{\mu}$ are:
$$
D^{\mu} A_{\mu \nu}- \frac{\mu^2}{1+\alpha^2} ( A_{\nu} + \alpha B_{\nu}) = 
i g \overline{\psi} \gamma _{\nu} T^i \psi T^i
\eqno{(6.2a)}
$$
$$
D_b^{\mu} B_{\mu \nu}- \frac{\alpha \mu^2}{1+\alpha^2} ( A_{\nu} + \alpha B_{\nu}) = 0 
\eqno{(6.2b)}
$$
respectively. In the above relations, we have used two simplified notations:
$$
D^{\mu} A_{\mu \nu} = \lbrack D^{\mu} ~,~ A_{\mu \nu} \rbrack
\eqno{(6.3a)}
$$
$$
D_b^{\mu} B_{\mu \nu} = \lbrack D_b^{\mu} ~,~ B_{\mu \nu} \rbrack . 
\eqno{(6.3b)}
$$

~~~~ Eqs(6.2a,b) can be expressed in terms of component fields $A^i_{\mu}$ and $B^i_{\mu}$:
$$
\partial ^{\mu} A^i_{\mu \nu}- \frac{\mu^2}{1+\alpha^2} ( A^i_{\nu} + \alpha B^i_{\nu}) = 
i g \overline{\psi} \gamma _{\nu} T^i \psi + g f^{ijk} A^j_{\mu \nu} A^{k \nu}
\eqno{(6.4a)}
$$
$$
\partial ^{\mu} B^i_{\mu \nu}- \frac{\alpha \mu^2}{1+\alpha^2} ( A^i_{\nu} + \alpha B^i_{\nu}) 
= - \alpha  g f^{ijk} B^j_{\mu \nu} B^{k \nu}
\eqno{(6.4b)}
$$
The equations of motion for gauge fields $C_{\mu}$ and $F_{\mu}$ can be easily obtained from 
eqs(6.4a,b). In other words, cos$\theta \cdot$ (6.4a) -- sin$\theta  \cdot$(6.4b) gives the equation of 
motion for massive vector field $C_{\mu}$, and  --sin$\theta \cdot$ (6.4a) + cos$\theta  
\cdot$(6.4b) gives the equation of motion for massless vector field $F_{\mu}$.  \\

~~~~ Please note that the above equations of motion are quite different from those of Yang-Mills 
gauge theory. But if $\alpha$ is small enough, two gauge theories will give similar results. Let
$$
\alpha  \ll  1 ,
\eqno{(6.5)}
$$
then, in the leading term, 
$$
{\rm cos} \theta \approx 1  ~~,~~ {\rm sin} \theta \approx 0 ,
\eqno{(6.6)}
$$
$$
A_{\mu} \approx C_{\mu} ~~,~~ B_{\mu} \approx F_{\mu}
\eqno{(6.7)}
$$
In this case, eqs(6.1) and (6.2a,b) change into
$$
\lbrack \gamma ^{\mu} ( \partial _{\mu} - i g   C_{\mu} ) +m \rbrack  \psi = 0 
\eqno{(6.8)}
$$
$$
D^{\mu} C_{\mu \nu}- \mu^2  C_{\nu}   = i g \overline{\psi} \gamma _{\nu} T^i \psi T^i
\eqno{(6.9)}
$$
$$
D_b^{\mu} F_{\mu \nu } = 0 
\eqno{(6.10)}
$$
respectively. Except for a mass term in eq(6.9), eqs(6.8-9) are the same as those in Yang-Mills 
gauge theory.  \\

~~~~ From eq(6.2a) or (6.2b), we can obtain a supplementary condition. Using eq(6.1), we can 
prove that 
$$
\lbrack D^{\lambda} ~,~ - i g \overline{\psi} \gamma _{\lambda} T^i \psi T^i \rbrack = 0.
\eqno{(6.11)}
$$
Let $D^{\nu}$ act on eq(6.2a) from the left, and let $D_b^{\nu}$ act on eq(6.2b) from the left, 
applying eq(5.5) and the following two identities: 
$$
\lbrack D^{\lambda} ~,~ \lbrack D^{\nu} ~,~A_{\nu \lambda} \rbrack \rbrack = 0
\eqno{(6.12a)}
$$
$$
\lbrack D_b^{\lambda} ~,~ \lbrack D_b^{\nu} ~,~B_{\nu \lambda} \rbrack \rbrack = 0 ,
\eqno{(6.12b)}
$$
we could obtain the following two equations
$$
\lbrack D^{\nu} ~,~  A_{\nu } + \alpha B_{\nu} \rbrack  = 0
\eqno{(6.13a)}
$$
$$
\lbrack D_b^{\nu} ~,~  A_{\nu } + \alpha B_{\nu} \rbrack  = 0
\eqno{(6.13b)}
$$
respectively. These two equations are essentially the same, they give a supplementary condition. If 
we expressed eqs(6.13a,b) in terms of component fields, these two equations will give the same 
expression:
$$
\partial ^{\nu} ( A^i_{\nu} + \alpha B^i_{\nu} ) + \alpha g f^{ijk} A^j_{\nu} B^{k \nu} = 0 .
\eqno{(6.14)}
$$

~~~~ When $\nu = 0$, eqs(6.4a,b) don't give dynamical equations of motion for gauge fields, 
because they contain no time derivative terms. They are just constrains. Originally, gauge fields 
$A^i_{\mu}$ and $B^i_{\mu}$ have $8(N^2-1)$ degrees of freedom, but they satisfy $2(N^2-
1)$constrains and have $ (N^2-1)$ gauge degrees of freedom, therefore, gauge fields $A^i_{\mu}$ 
and $B^i_{\mu}$ have $5(N^2-1)$ independent dynamical degrees of freedom altogether. This 
result coincides with our experience: a massive vector field has 3 independent degrees of freedom and 
a massless vector field has 2 independent degrees of freedom. \\

\section{The Case That Matter Fields Are Scalar Fields}

~~~~ In the above discussions, all matter fields are spinor fields. Now, we consider the case when 
matter fields are scalar fields. Suppose that there are $N$ scalar fields $\varphi _l (x) ~(l=1,2, \cdots 
N)$ which form a multiplet of matter fields:
$$
\varphi (x) =\left ( 
\begin{array}{c}
\varphi_1 (x) \\
\varphi_2 (x) \\
\vdots \\
\varphi_N (x)
\end{array}
\right ) 
\eqno{(7.1)} 
$$
All $\varphi (x)$ form a representative space of $SU(N)$ group. In gauge transformation, $\varphi 
(x)$ 
transforms as :
$$
\varphi (x) \longrightarrow \varphi ' (x) = U \varphi (x)
\eqno{(7.2)}
$$

~~~~ The lagrangian density is 
$$
\begin{array}{ccl}
\cal L &= &- \lbrack (\partial _{\mu} - i g A_{\mu} ) \varphi \rbrack ^{+}
(\partial ^{\mu} - i g A^{\mu} ) \varphi - V( \varphi ) \\
&&-\frac{1}{4K} Tr( A^{\mu \nu} A_{\mu \nu} )
-\frac{1}{4K} Tr( B^{\mu \nu} B_{\mu \nu} ) \\
&&-\frac{\mu ^2}{2K ( 1+ \alpha ^2)} 
Tr \left \lbrack (A^{\mu}+\alpha B^{\mu})( A_{\mu}+\alpha B_{\mu} ) 
\right \rbrack
\end{array}
\eqno{(7.3)} 
$$
The above lagrangian density can be expressed in terms of component fields :
$$
\begin{array}{ccl}
\cal L &= &- \lbrack (\partial _{\mu} - i g A^i_{\mu} T_i ) \varphi \rbrack ^{+}
(\partial ^{\mu} - i g A^{i \mu} T_i ) \varphi - V( \varphi ) \\
&&-\frac{1}{4}  A^{i \mu \nu} A^i_{\mu \nu} 
-\frac{1}{4} B^{i \mu \nu} B^i_{\mu \nu}  \\
&&-\frac{\mu ^2}{2 ( 1+ \alpha ^2)} 
 (A^{i \mu}+\alpha B^{i \mu})( A^i_{\mu}+\alpha B^i_{\mu} ) 
\end{array}
\eqno{(7.4)} 
$$
The general form for $V(\varphi)$ which is renormalizable and gauge invariant is
$$
V(\varphi) = m^2 \varphi^{+} \varphi + \lambda (\varphi ^{+} \varphi)^2 .
\eqno{(7.5)}
$$
It is easy to prove that the lagrangian density ${\cal L}$ defined by eq(7.3) has local $SU(N)$ gauge 
symmetry. The Euler-Lagrange equation of motion for scalar field $\varphi$ is:
$$
 (\partial ^{\mu} - i g A^{\mu} ) (\partial _{\mu} - i g A_{\mu} ) \varphi - m^2 \varphi 
-  2 \lambda \varphi ( \varphi^{+} \varphi )^2=0
\eqno{(7.6)}
$$

~~~~ If $N^2-1$ scalar fields $\varphi _l (x) ~(l=1,2, \cdots N^2-1)$ form a multiplet of matter 
fields
$$
\varphi (x) = \varphi _l (x) T_l ,
\eqno{(7.7)}
$$
then, the gauge transformation of $\varphi(x)$ should be
$$
\varphi (x) \longrightarrow \varphi ' (x) = U \varphi (x) U^{+} .
\eqno{(7.8)}
$$
All $\varphi (x)$ form a space of adjiont representation of $SU(N)$ group. In this case, the gauge 
covariant derivative is
$$
D_{\mu} \varphi = \partial _{\mu} \varphi - i g \lbrack A_{\mu} ~,~ \varphi \rbrack , 
\eqno{(7.9)}
$$
and the gauge invariant lagrangian density ${\cal L}$ is 
$$
\begin{array}{ccl}
\cal L &= &- \frac{1}{K} Tr \lbrack (D^{\mu} \varphi) ^{+}(D_{\mu} \varphi) \rbrack - 
V( \varphi ) \\
&&-\frac{1}{4}  A^{i \mu \nu} A^i_{\mu \nu} 
-\frac{1}{4} B^{i \mu \nu} B^i_{\mu \nu}  \\
&&-\frac{\mu ^2}{2 ( 1+ \alpha ^2)} 
 (A^{i \mu}+\alpha B^{i \mu})( A^i_{\mu}+\alpha B^i_{\mu} ) .
\end{array}
\eqno{(7.10)} 
$$

\section{A More General Model}

~~~~ In the above discussions, we have constructed a gauge field model which has rigorous 
$SU(N)$ gauge symmetry and massive gauge bosons. In the above model, only gauge field 
$A_{\mu}$ interacts with matter fields $\psi$ or $\varphi$, gauge field $B_{\mu}$ doesn't interact 
with matter fields. Now, we will construct a more general gauge field model, in which both gauge 
fields interact with matter fields. And in a proper limit, this model will return to the above model. As 
an example, we only discuss the case when matter fields are spinor fields. The case when matter 
fields are scalar fields can be discussed similarly.  \\

~~~~ In chapter 4, we have prove that, under local gauge transformations, $D_{\mu}$ and $D_{b 
\mu}$ transform covariantly. It is easy to prove that ${\rm cos}^2 \phi D_{\mu} + {\rm sin}^2 \phi 
D_{b \mu}$ is the most general gauge covariant derivative which transforms covariantly under local 
$SU(N)$ gauge transformations
$$
{\rm cos}^2 \phi D_{\mu} + {\rm sin}^2 \phi D_{b \mu} \longrightarrow  
U({\rm cos}^2 \phi D_{\mu} + {\rm sin}^2 \phi D_{b \mu} )U^{+}, 
\eqno{(8.1)}
$$
where $\phi$ is constant. Then the following lagrangian has local $SU(N)$ gauge symmetry
$$
\begin{array}{ccl}
\cal L &= &- \overline{\psi} \lbrack \gamma ^{\mu} ({\rm cos}^2 \phi D_{\mu} 
+{\rm sin}^2 \phi D_{b \mu}) +m \rbrack \psi  \\
&&-\frac{1}{4K} Tr( A^{\mu \nu} A_{\mu \nu} )
-\frac{1}{4K} Tr( B^{\mu \nu} B_{\mu \nu} ) \\
&&-\frac{\mu ^2}{2K ( 1+ \alpha ^2)} 
Tr \left \lbrack (A^{\mu}+\alpha B^{\mu})( A_{\mu}+\alpha B_{\mu} ) 
\right \rbrack
\end{array}
\eqno{(8.2)} 
$$
Let ${\cal L}_{\psi}$ denote the part for fermions:
$$
{\cal L}_{\psi} = - \overline{\psi} \lbrack \gamma ^{\mu} ({\rm cos}^2 \phi D_{\mu} 
+{\rm sin}^2 \phi D_{b \mu}) +m \rbrack \psi.  
\eqno{(8.3)}
$$
Using eqs(2.9a,b), we can change ${\cal L}_{\psi}$ into 
$$
{\cal L}_{\psi} = - \overline{\psi} \lbrack \gamma ^{\mu} ( \partial _{\mu} 
- i g {\rm cos}^2 \phi A_{\mu} + i \alpha g {\rm sin}^2 \phi B_{ \mu}) +m \rbrack \psi . 
\eqno{(8.4)}
$$
From the above lagrangian, we know that both gauge fields $A_{\mu}$ and $B_{\mu}$ couple with 
matter field $\psi$. Substitute eqs(5.8a,b) into eq(8.4), we get
$$
{\cal L}_{\psi} = - \overline{\psi} \lbrack \gamma ^{\mu} ( \partial _{\mu} 
- i g  \frac{{\rm cos}^2 \theta -{\rm sin}^2 \phi}{{\rm cos }\theta}  C_{\mu} 
+ i g {\rm sin}\theta F_{ \mu}) +m \rbrack \psi . 
\eqno{(8.5)}
$$

~~~~ The equation of motion for fermion field $\psi$ is
$$
\lbrack \gamma ^{\mu} ( \partial _{\mu} 
- i g  \frac{{\rm cos}^2 \theta -{\rm sin}^2 \phi}{{\rm cos }\theta}  C_{\mu} 
+ i g {\rm sin}\theta F_{ \mu}) +m \rbrack \psi = 0 . 
\eqno{(8.6)}
$$
The equations of motion for gauge fields $A_{\mu}$ and $B_{\mu}$  now change into:
$$
D^{\mu} A_{\mu \nu}- \frac{\mu^2}{1+\alpha^2} ( A_{\nu} + \alpha B_{\nu}) = 
i g {\rm cos}^2 \phi  \overline{\psi} \gamma _{\nu} T^i \psi T^i
\eqno{(8.7a)}
$$
$$
D_b^{\mu} B_{\mu \nu}- \frac{\alpha \mu^2}{1+\alpha^2} ( A_{\nu} + \alpha B_{\nu}) = 
- i \alpha g {\rm sin}^2 \phi  \overline{\psi} \gamma _{\nu} T^i \psi T^i .
\eqno{(8.7b)}
$$

~~~~ If $\phi$ vanish, the lagrangian density (8.2) will become the original lagrangian density 
(2.13), 
the equations of motion (8.7a,b) will return to eqs(6.2a,b), and eq(8.6) will return to eq(6.1). So, the 
model discussed in the above chapters is just a special case of the model we discuss now. \\

\section{U(1) Case}

~~~~ If the symmetry of the model is U(1) group, we will obtain a U(1) gauge field model. We also 
use $A_{\mu}$ and $B_{\mu}$  to denote gauge fields and $\psi$ to denote a multiplet of fermion 
fields. In U(1) case, the strengths of gauge fields are
$$
A_{\mu \nu}  =  \partial _{\mu} A_{\nu} - \partial _{\nu} A_{\mu}
\eqno{(9.1a)} 
$$
$$
B_{\mu \nu}  =  \partial _{\mu} B_{\nu} - \partial _{\nu} B_{\mu}
\eqno{(9.1b)} 
$$
Two gauge covariant derivatives are the same as (2.9a,b) but with different content. The lagrangian 
density of the model is:
$$
\begin{array}{ccl}
\cal L &= &- \overline{\psi} \lbrack \gamma ^{\mu} ({\rm cos}^2 \phi D_{\mu} 
+{\rm sin}^2 \phi D_{b \mu}) +m \rbrack \psi  \\
&&-\frac{1}{4}  A^{\mu \nu} A_{\mu \nu} 
-\frac{1}{4}  B^{\mu \nu} B_{\mu \nu}  \\
&&-\frac{\mu ^2}{2 ( 1+ \alpha ^2)} 
 (A^{\mu}+\alpha B^{\mu})( A_{\mu}+\alpha B_{\mu} ) 
\end{array}
\eqno{(9.2)} 
$$

~~~~ The local U(1) gauge transformations are
$$
\psi \longrightarrow e^{- i \theta} \psi ,
\eqno{(9.3a)} 
$$
$$
A_{\mu} \longrightarrow  A_{\mu} -\frac{1}{g} \partial _{\mu}\theta
\eqno{(9.3b)} 
$$
$$
B_{\mu} \longrightarrow  B_{\mu} +\frac{1}{ \alpha g} \partial _{\mu}\theta .
\eqno{(9.3c)} 
$$
Then, $A_{\mu \nu} ,~B_{\mu \nu}$ and $A_{\mu}+\alpha B_{\mu}$ are all U(1) gauge invariant. 
That is
$$
A_{\mu \nu} \longrightarrow  A_{\mu \nu}
\eqno{(9.4a)} 
$$
$$
B_{\mu \nu} \longrightarrow  B_{\mu \nu}
\eqno{(9.4b)} 
$$
$$
A_{\mu}+\alpha B_{\mu} \longrightarrow A_{\mu}+\alpha B_{\mu}
\eqno{(9.4c)} 
$$
Using all these results, it is easy to prove that the lagrangian density given by eq(9.2) has local 
$U(1)$ gauge symmetry.  \\

~~~~ Substitute eqs(5.8a,b) into eq(9.2), the lagrangian density ${\cal L}$ changes into
$$
\begin{array}{ccl}
{\cal L} & = & - \overline{\psi} \lbrack \gamma ^{\mu} ( \partial _{\mu} 
- i g  \frac{{\rm cos}^2 \theta -{\rm sin}^2 \phi}{{\rm cos }\theta}  C_{\mu} 
+ i g {\rm sin}\theta F_{ \mu}) +m \rbrack \psi \\ 
&&-\frac{1}{4}  C^{\mu \nu} C_{\mu \nu} 
-\frac{1}{4}  F^{\mu \nu} F_{\mu \nu} - \frac{\mu ^2}{2} C^{\mu} C_{\mu}
\end{array}
\eqno{(9.5)} 
$$
Now, we see that there is a massive Abel gauge field as well as a massless Abel gauge field. They all 
have gauge interaction with matter field. In this case, $U(1)$ gauge interaction is transmitted by two 
different kinds of gauge fields.\\

\section{Yang-Mills Limit}

~~~~ In chapter six, we have discussed the Yang-Mills limit of equations of motion. Now, starting 
from the lagrangian density of the model, we will discuss the Yang-Mills limit of the model. There 
are two kinds of Ynag-Mills limit of the present gauge field theory. \\

~~~~ The first kind of Yang-Mills limit corresponds to very small parameter $\alpha$. Let
$$
\alpha \longrightarrow 0 ,
\eqno{(10.1)}
$$
then
$$
{\rm cos} \theta = 1 ~~,~~ {\rm sin}\theta =0.
\eqno{(10.2)}
$$
From eqs(5.7a,b), we know that the gauge field $A_{\mu}$ is just gauge field $C_{\mu}$ and the 
gauge field $B_{\mu}$ is just the gauge field $F_{\mu}$. That is
$$
C_{\mu} = A_{\mu} ~~,~~ F_{\mu}=B_{\mu} .
\eqno{(10.3)}
$$
In this case, the lagrangian density (2.14) becomes
$$
\begin{array}{ccl}
{\cal L} & = & - \overline{\psi} \lbrack \gamma ^{\mu} ( \partial _{\mu} 
- i g  C^i_{\mu} T^i ) +m \rbrack \psi \\ 
&&-\frac{1}{4}  C^{i \mu \nu} C^i_{\mu \nu} 
-\frac{1}{4}  F^{i \mu \nu} F^i_{\mu \nu} - \frac{\mu ^2}{2} C^{i \mu} C^i_{\mu} .
\end{array}
\eqno{(10.4)} 
$$
Please note that massless gauge fields do not interact with matter fields. So, the  $\alpha 
\longrightarrow 0 $ limit corresponds to the case that gauge interaction is mainly transmitted by 
massive gauge fields. In other words, the above lagrangian describe those gauge interactions of 
which the masses of intermediate bosons are non-zero. It is known that electroweak interactions 
belong to this category. Except for a mass term of gauge fields, the above lagrangian density is the 
same  as that of the Yang-Mills theory. But if $\alpha$ strictly vanishes, the lagrangian does not 
have gauge symmetry and the theory is not renormalizable.  \\

~~~~ The second kind of Yang-Mills limit corresponds to very big parameter $\alpha$. Let
$$
\alpha \longrightarrow \infty ,
\eqno{(10.5)}
$$
then
$$
{\rm cos} \theta = 0 ~~,~~ {\rm sin}\theta =1.
\eqno{(10.6)}
$$
From eqs(5.7a,b), we know that the gauge field $B_{\mu}$ is just the gauge field $C_{\mu}$ and 
the gauge field $A_{\mu}$ is just the gauge field $-F_{\mu}$. That is
$$
C_{\mu} = B_{\mu} ~~,~~ F_{\mu}= - A_{\mu} .
\eqno{(10.7)}
$$
Then, the lagrangian density (2.14) becomes
$$
\begin{array}{ccl}
{\cal L} & = & - \overline{\psi} \lbrack \gamma ^{\mu} ( \partial _{\mu} 
+ i g  F^i_{\mu} T^i ) +m \rbrack \psi \\ 
&&-\frac{1}{4}  F^{i \mu \nu} F^i_{\mu \nu} 
-\frac{1}{4}  C^{i \mu \nu} C^i_{\mu \nu} - \frac{\mu ^2}{2} C^{i \mu} C^i_{\mu} .
\end{array}
\eqno{(10.8)} 
$$
In this case, massive gauge fields do not interact with matter fields. So, this limit corresponds to the 
case when gauge interaction is mainly transmitted by massless gauge fields. Similarly, the 
lagrangian density (10.8) do not have strict gauge symmetry.  \\

~~~~ In the model of particles' interaction which describes the gauge interaction of real world, the 
parameter $\alpha$ should be finite,
$$
0 < \alpha < \infty.
\eqno{(10.9)}
$$
In this case, both massive gauge fields and massless gauge fields interact with matter fields, and 
gauge interaction is transmitted by both of them. \\

\section{Discussion}

~~~~ In chapter six, we have said that when $\alpha \ll 1$, the equations of motion given by gauge 
fields model discussed in this paper are similar to those of Yang-Mills gauge theory except for a mass 
term. So, we could anticipate that these two gauge theories will give similar dynamical behaviors in 
describing particles' interaction. \\

~~~~ If we apply this model to strong interaction  \lbrack 14 \rbrack, we will obtain two sets of 
gluons: one set is 
massive and another set is massless. Because of color confinement, all colored gluons are confined. 
The important thing is that there may exist three sets of glueballs. If we apply this model to 
electroweak interactions, we will obtain two sets of intermediate gauge bosons: one set is massive 
which has already been found by experiment and another set is massless. \lbrack 15 \rbrack. In this 
new electroweak model, there is no Higgs particle. Because the parameter $\alpha$ is unknown, there 
exists no contradiction between the theory discussed in this paper and experiment.   \\

~~~~ The new theory predicts many new massless , electric neutral vector particles. In the high 
energy experiment, it is hard to distinguish between all these massless, electric neutral vector 
particles and $\gamma$ photon. Experimental physicists have found that $\gamma$ photon takes 
part 
in strong interaction and electroweak interactions \lbrack 16 \rbrack. This phenomenon means that 
there are some massless, electric neutral vector particles mixed in $\gamma$ photon.  \\

~~~~ Because the model has strict gauge symmetry, we can anticipate that this gauge field theory is 
renormalizable \lbrack 17 \rbrack . \\

\section*{Reference:}
\begin{description}
\item[\lbrack 1 \rbrack]  C.N.Yang, R.L.Mills, Phys Rev {\bf 96} (1954) 191
\item[\lbrack 2 \rbrack]  S.Glashow, Nucl Phys {\bf 22}(1961) 579
\item[\lbrack 3 \rbrack]  S.Weinberg, Phys Rev Lett {\bf 19} (1967) 1264
\item[\lbrack 4 \rbrack]  A.Salam, in Elementary Particle Theory, eds.N.Svartholm(Almquist and 
Forlag, Stockholm,1968)
\item[\lbrack 5 \rbrack]  T.D.Lee, M Rosenbluth, C.N.Yang, Phys. Rev. {\bf 75} (1949) 9905
\item[\lbrack 6 \rbrack]  J. Goldstone, Nov. Cim. {\bf 19} (1961) 154
\item[\lbrack 7 \rbrack]  Y Nambu, G.Jona-Lasinio, Phys.Rev. {\bf 122} (1961)345
\item[\lbrack 8 \rbrack]  J.Goldstone, A.Salam, S.Weinberg.  Phys.Rev.{\bf 127}(1962)965
\item[\lbrack 9 \rbrack]  P.W.Higgs, Phys.Lett.{\bf 12}(1964) 132
\item[\lbrack 10 \rbrack]  F.Englert, R.Brout Phys.Rev.Lett. {\bf 13} (1964) 321
\item[\lbrack 11 \rbrack]  G.S.Guralnik, C.R.Hagen, T.W.B.Kibble  Phys.Rev.Lett. {\bf 13} 
(1964) 585
\item[\lbrack 12 \rbrack]  P.W.Higgs Phys.Rev. {\bf 145} (1966) 1156
\item[\lbrack 13 \rbrack]  S.Weinberg, Phys.Rev. {\bf D7} (1973) 1068
\item[\lbrack 14 \rbrack]  Ning Wu, A new model for strong interaction
(hep-ph/9802297)
\item[\lbrack 15 \rbrack]  Ning Wu, A new model for electroweak interactions
(hep-ph/9802237)
\item[\lbrack 16 \rbrack] There are many papers discuss the photon interaction experimentally. See 
for example hep-ex /9711005
\item[\lbrack 17 \rbrack]  Ning Wu, The Renormalization of the Non-Abel Gauge Field Theory  
 with massive gauge bosons (in preparation)
\end{description}

\end{document}